\documentstyle[12pt,epsf]{article}
\newcommand{\cftnote}
{\renewcommand{\thefootnote}{\fnsymbol{footnote}}}
\newcommand{\resetftnote}{\setcounter{footnote}{0}}

\newcommand{\be}{\begin{equation}}
\newcommand{\ee}{\end{equation}}
\newcommand{\bea}{\begin{eqnarray}}
\newcommand{\eea}{\end{eqnarray}}

\def\bm#1{\mbox{\boldmath{$#1$}}}

\begin{document}

\begin{flushright}
ITFA-94-31, hep-th/9412191\\
November 1994
\end{flushright}

\cftnote

\vspace{1mm}
{\Large \bf
\begin{center}
Deformed 2d CFT:\\ Landau-Ginzburg Lagrangians and Toda \nolinebreak
theories%
\footnote{Research supported in part by CICYT (Spain) and
FOM (The Netherlands).}
\end{center}
}
\vspace{2mm}
\setcounter{footnote}{3}
\begin{center}
{\large Jos\'e Gaite%
\footnote{E-mail address: iffgaite@roca.csic.es}}
\\[4mm]
{\it Instituut voor Theoretische Fysica, University of Amsterdam,\\
Valckenierstraat 65, 1018 XE Amsterdam, The Netherlands}\\[2mm] and \\[2mm]
{\it Instituto de Matem\'aticas y
F\'{\i}sica Fundamental, C.S.I.C.,\\ Serrano 123, 28006 Madrid,
Spain.}\\
\end{center}
\resetftnote

\vspace{2mm}

\begin{abstract}
We consider the relation between affine Toda field theories (ATFT) and
Landau-Ginzburg Lagrangians as alternative descriptions of
deformed 2d CFT. First, we show that the two concrete implementations
of the deformation are consistent once quantum corrections to the
Landau-Ginzburg Lagrangian are taken into account.
Second, inspired by Gepner's fusion potentials,
we explore the possibility of a direct connection between both types
of Lagrangians; namely, whether they can be transformed one into another
by a change of variables.
This direct connection exists in the one-variable case, namely,
for the sine-Gordon model, but cannot be established in general.
Nevertheless, we show that both potentials exhibit
the same structure of extrema.
\end{abstract}

\global\parskip 4pt

\section{Introduction}

The study of 2d FT in the vicinity of a critical point has gained importance
over the latter years. There are essentially two ways of investigation:
The first is to analyse the critical behavior of 2d statistical models,
in particular, of integrable type. The second begins with the critical model,
some 2d CFT, and deforms it in a suitable manner such that the correlation
length becomes finite. The most interesting deformations are
again the so-called
integrable, for which the non-critical model still has an infinite number of
conservation laws that allow to calculate correlation functions. The simplest
example is the thermal perturbation of the Ising model. It generalizes to the
perturbation of minimal models of the Virasoro algebra by
the most relevant field, $\Phi_{(1,3)}$.
It further generalizes to the perturbation of minimal models of W-algebras
by the most relevant field of the thermal subalgebra,
$\Phi(1, \cdots, 1 \mid {\rm Adj})$.
The integrability of these perturbations lies on the fact that the deformed
theories can be realized as quantum affine Toda field theories (ATFT)
with imaginary coupling constant \cite{Mans,EY}.

Affine Toda field theories are defined in terms of a classical
Lagrangian with a potential of exponential form.
The coupling constant $\beta$
appears in the exponential; hence, when it takes imaginary values then
the potential is periodic in
the fundamental field and can therefore have soliton solutions.%
\footnote{See \cite{Hollo} for a study of these solitons.}
The description of quantum affine Toda field theories is best attained
in terms of these solitons;
due to integrability, their only interaction consists of elastic scattering.
One can say that the structure of degenerate ground states of the
classical potential is the dominant feature of these models
for it determines the soliton types.
The series $W_{(n)}^{\,p}$ of minimal models
are obtained for $\beta^2 = -p/p+1$. These are called
restricted theories since the soliton spectrum has to be restricted
for consistency. Then only a finite number of ground states
and therefore a portion of the potential are allowed.

There is another Lagrangian description of
deformed 2d CFT with W-symmetry, namely,
the one given by Landau-Ginzburg (LG) Lagrangians.
This description was introduced in a previous paper \cite{I}.
There was shown as well that those Landau-Ginzburg Lagrangians
correspond to the universality classes of critical behavior of the
integrable models of Jimbo et al \cite{Jimbo}.
To be precise, it was shown that there exists a perturbation of the
Landau potential such that its ground states exactly correspond to
the ground states of regime III of Jimbo et al models.
Hence, this perturbed Landau potential possesses
features similar to those of the portion of potential
that characterizes restricted ATFT.

The question arises of whether a direct connection between these two
types of classical descriptions of deformed 2d CFT can be established.
The first problem that comes to one's mind is that the deformations
that produce the mentioned ground state structure seem to be
different in either approach. In the case of the affine Toda field theory,
it is $\Phi(1, \cdots, 1 \mid {\rm Adj})$ whereas in the case of
the Landau-Ginzburg theory not only the composite field
corresponding to it intervenes but
a combination of all the symmetry
preserving relevant fields.
This difference is due to the classical character of Landau-Ginzburg
Lagrangians. The perturbed Lagrangian must include quantum corrections,
bringing about all the other symmetry preserving relevant fields.
We study this effect in the first section.
We argue that for $n=2$ the quantum effective potential is directly
related to that of the corresponding ATFT, the sine-Gordon potential.
In the second section we formalize the correspondence between
$n=2$ ATFT and the restricted sine-Gordon model.
We generalize to other $n$ using Gepner's ideas \cite{Gepner}.
Finally, we study the ground states of affine Toda
potentials and we compare them with those given by Landau potentials.

The connection between ATFT and Landau-Ginzburg Lagrangians for
CFT with $N=2$ supersymmetry has been studied in \cite{Warner}.
Since these Landau-Ginzburg Lagrangians are not related to those
of the bosonic theories, there is no apparent relation with
our work.

\section{Quantum corrections to Landau-Ginzburg Lagrangians}

Let us introduce some terminology of thermodynamics to better describe
the features of the phase diagrams in which we are interested.
The ground states of regime III of Jimbo et al models are
ordered phases that coexist along a manifold ending at the multicritical
point. We are interested in the particular line of
multiple coexistence given by
perturbation with $\Phi(1, \cdots, 1 \mid {\rm Adj})$.
It is not difficult to find the form of the equation for
multiple coexistence in terms of the coefficients of the Landau potential.
To be definite, let us consider the case with
one order paramenter $\varphi$. The potential $V = \varphi^{2\,k}$ has
$k$ minima which unfold under perturbation. A generic perturbation is given
by a polynomial in $\varphi$ of degree $2\,k - 2$, which is thus
the codimension of the multicritical point. Since the heights of the minima
with respect to a given one are specified by $k-1$ numbers, this is the
codimension of the multiple coexistence manifold. If we only admit
$Z_2$ symmetric perturbations, the codimension of the multicritical point
is $k-1$ and the codimension of the multiple coexistence manifold
$(k-1)/2$ or $(k/2)-1$ for $k$ odd or even, respectively.

For example,
let $k = 4$. The perturbed symmetric potential is
\be
V = \varphi^8 + u\,\varphi^6 + v\,\varphi^4 + w\,\varphi^2.
\ee
It can also be written as
\be
V = (\varphi^2 - b_1^2)^2 \left((\varphi^2 - b_2^2)^2 + d\right)
\ee
with another set ot parameters more related to its shape.
The condition for quadruple coexistence is
$d = 0$. Then the minima
occur for $\varphi = \pm b_1, \pm b_2$.
That condition is expressed in $uvw$-space as a surface
(codimension-one submanifold) with parametric equations
\bea
u&=& -2\, (b_1^2+b_2^2), \\
v &=& b_1^4 + 4\, b_1^2\,b_2^2 + b_2^4,   \\
w &=& -2\, (b_1^2\,b_2 + b_2^2\,b_1).
\eea
Since the position of the minima with positive coordinates, $b_1$ and $b_2$,
is still arbitrary, we can impose one further condition on their ratio,
$b_2 = \alpha\, b_1$,
thus fixing the shape of the potential up to overall rescaling.
This restricted multiple coexistence occurs along lines which end at the
tetracritical point. Their parametric dependence is given by dimensional
analysis, while the coefficients, as funtions of $\alpha$,
specify a particular line. The important point is that $v$ and $w$ take
non-null values along these lines. The line $v = w = 0$ and $u \leq 0$
is not of coexistence but a critical metastable line on which the two middle
minima disappear and is not of interest.%
\footnote{It actually belongs to the critical manifold that separates
regimes III and IV.}
In contrast, we know that the exact quantum perturbation that produces
multiple coexistence is $\Phi_{(1, 3)}$, corresponding to
$\varphi^{2\,k-2}$; hence only $u$ being non-null.
However, the identification of primary fields with composite fields
in the Landau-Ginzburg approach is only valid at the critical point.
In the perturbed theory there are quantum effects
that induce field mixing.
The least relevant perturbation $\varphi^{2\,k-2}$ has the coupling
of lowest dimension (in space dimension $d>2$). Therefore,
a perturbation with it or, in other words, a non-null value of
its coupling constant will contribute
to all other coupling constants, of higher dimension.

To analyse quantum effects, let us consider the simplest case, $k=3$,
with potential
\be
V = u\,\varphi^6 + v\,\varphi^4 + w\,\varphi^2.
\ee
At second order perturbation theory in $v$ there is a contribution
to the two-point vertex, $w=c\,v^2$,
given by the setting-sun Feynman diagram
\be
\setlength{\unitlength}{0.75cm}
\begin{picture}(2,1)
\put(0,0.5){\line(1,0){2}}
\put(1,0.5){\circle{1}}
\end{picture}.                    \label{diag1}
\ee
The value of the coupling $u$ can be taken to be its fixed-point
value $u^*$. It is obtained as the non-trivial
solution of $\beta(u)=0$. The $\beta$-function is calculated from
the 6-point vertex, given at second order by the diagram
\be
\setlength{\unitlength}{0.75cm}
\begin{picture}(2,1)
\put(0,0.5){\line(1,0){2}}
\put(0.5,0.5){\line(-1,1){0.5}}
\put(0.5,0.5){\line(-1,-1){0.5}}
\put(1.5,0.5){\line(1,1){0.5}}
\put(1.5,0.5){\line(1,-1){0.5}}
\put(1,0.5){\circle{1}}
\end{picture}\;.                         \label{diag2}
\ee
Upon calculation of these diagrams one obtains $c$ and $u^*$.
It is convenient to normalize the potential as before,
dividing by $u^*$ and redefining the two relevant couplings,
$\tilde{v}=v/u^*$ and $\tilde{w}=w/u^*$. They are related through
\be
\tilde{w}={c\, u^*}\,\tilde{v}^2.      \label{per-tl}
\ee
We have to compare this relation
with the equation of the triple line.
This equation can be simply obtained by writing the potential as
\be
V = \left(\varphi^2 - b^2\right)^2 \varphi^2
\ee
in the parametric form
\bea
v &=& -2\,b^2,  \\
w &=& b^4
\eea
or
\be
w - {1 \over 4}\,v^2 = 0.                \label{tl}
\ee
It has the same form as (\ref{per-tl}). To compare the coefficients
of $v^2$ we have to evaluate ${c\, u^*}$. Fortunately, we do not
need to evaluate both diagrams independently but its quotient,
since $u^*$ is in inverse proportion to the value of (\ref{diag2}).
Furthermore, we note that both diagrams are essentially the
same, namely, the propagator cubed ($\Delta^3$),
except for combinatorial factors. These factors can be found
for any $k$ in \cite{ID-HW}. Thus we obtain the
{\em dimension-independent} value
\be
c\, u^* = \frac{{(4!)^2\over 6}\,\Delta^3}
{{(6!)^2\over{2(3!)^3}}\,\Delta^3} = 0.08.
\ee
It is substantially smaller than 1/4 in (\ref{tl}). The difference is
due to higher order corrections,
which are known to be non-negligible \cite{ID-HW}.

A similar argument can me made in favor of the fact
that a multiple coexistence
line calculated in perturbation theory within the Landau-Ginzburg approach
converges to the exact line provided by the
$\Phi(1, \cdots, 1 \mid {\rm Adj})$
perturbation. However, given the computational restrictions, we
must content ourselves by saying that
any W-algebra Landau-Ginzburg Lagrangian
perturbed by the least relevant field develops due to quantum effects
non-null values for the coupling constants of the other
(symmetry-preserving) relevant fields such that the renormalized
equation for that line is compatible with the exact one.

The problem that we face now is that, in general,
there is an entire manifold of coexistence, as we saw for $k=4$ before.
In absence of an accurate perturbative solution,
we have to consider further
conditions to define a unique line that corresponds to the ATFT.
The difference between coexistence lines can be assumed
to represent the shape of the potential while the parameter along lines
is a scale variable representing its size.
The essential feature of the shape of the potential
is the height of the walls that separate its ground states. This height
measures how close the respective phases are in a thermodynamical
sense. In field theory it determines the energy of the kinks
that interpolate between ground states. A reasonable condition
may be to take all those heights equal.
In so doing, the generic potential for the Virasoro series, with
one order parameter, takes a particularly simple form, namely,
of a Chebyshev polynomial of the first kind.%
\footnote{The theories with $N=2$ supersymmetry are
much more constrained,
to the extent that it is possible to establish recursion relations
among perturbed composite fields which can be solved to give
the Chebyshev polynomial potential \cite{DijVer2}.}
This type of polynomial is defined by
\be
T_n(\cos\phi) = \cos(n\,\phi),  \label{Cheby}
\ee
where the degree is $n = 2\,k$ in our case.
Obviously it has the set of maxima and the set of minima
at the same height, which is 1 or $-1$, respectively.
Furthermore, with $\phi$ as the fundamental field it is
the potential of the sine-Gordon model, the simplest ATFT.
In this light, we see that the condition of equal heights above
is, besides natural, the only one that leads to this correspondence.

The Chebyshev polynomial ansatz provides definite values for
the coupling constants of the quantum potential.
It also gives the field mixing structure. In particular,
the least relevant composite field, $\varphi^{2k-2}$,
becomes
\be
\left[\varphi^{2k-2}\right] \equiv {\partial V \over \partial u} =
\varphi^{2k-2} + \cdots.             \label{Cheby2}
\ee
An examination of the coefficients in this expression shows that
it can be identified as
Chebyshev polynomial of the second kind, $U_{2k-2}(\varphi)$.
The appearance of
this particular type of polynomial is not coincidental,
according to the ideas introduced in the next section.

The previous line of reasoning is in principle
applicable to the W-series, with potentials
with several order parameters. The effect of quantum corrections is
essentially the same. The definition of a unique line
of coexistence is however more delicate, since the top of
the wall which separates two minima may not correspond to a maximum but
to a saddle point. We can try, regardless of this problem, to generalize
the previous form in terms of Chebyshev polynomials. For example,
in the $W_3$ case it is convenient a triangular coordinate system which
consists of 3 coordinates, $x_1$, $x_2$ and $x_3$,
subject to the constraint $x_1 + x_2 + x_3 = 0$.
A simple potential with the correct symmetry
that generalizes the one-variable case is the sum
\be
V = \sum_{i=1}^3 T_{2k}(x_i).   \label{trial}
\ee
Although the polynomial degree is correct,
we do not achieve the potentials
obtained in \cite{I}. The first non-trivial case (3-state Potts model)
occurs when $k = 2$. Now, it is straightforward to verify that for $k=2$
the potential (\ref{trial}) has no third order term and hence it has
full rotational symmetry, contrary to the correct $D_3$ symmetry.%
\footnote{In general, the symmetry of (\ref{trial})
is $D_6$, as a consequence of
the absence of odd powers of the third order term,
nevertheless higher than
the required $D_3$ symmetry.}
A more subtle way to generalize
the Chebyshev polynomials to higher dimension is required.
We concern ourselves with this problem in the next section.

\section{Landau-Ginzburg Lagrangians as
restricted af\-fine Toda theories}

We have seen above that the Landau potential of
the Virasoro minimal models for the required
type of phase coexistence is closely related to the potential of
the sine-Gordon model. The latter has infinitely many ground states,
all placed periodically,
while the former has only a finite number. Thus the Landau potential
can only represent a strecht of the sine-Gordon model.
Besides, it is not a periodic function
but a polynomial. Of course, these differences come from the different
variables used to define them; they are related by
\be
\varphi \simeq \cos\phi.   \label{id}
\ee
Hence, the only relevant values of the sine-Gordon field
are $0 < \phi < \pi$, for which $-1 < \varphi < 1$.
The number of extrema
that fit in this interval is $2k$, the degree of the Chebyshev
polynomial (\ref{Cheby}).
The deep reason for the restriction in the number of minima is
the truncation of the soliton spectrum that occurs
at certain values of the coupling constant:
The soliton and antisoliton which this model
possess belong to a representation of a quantum
group. For certain representations ($q$ is a root of unity)
there is truncation and only
a finite number of solitons and antisolitons can be composed \cite{LeClair}.%
\footnote{Do not confuse this truncation in the spectrum with the
truncation in the OPE alluded to before.}
These solitons precisely correspond to the kinks of the Landau potential.
The details of this truncation are not essential in
the present context.

The picture based on the identification (\ref{id}) is
of a qualitative nature yet. The reader may have noticed that
our actual sine-Gordon potential is
\be
V(\varphi) = \cos\left(\beta\,\phi \right),    \label{sG}
\ee
with
\be
\beta^2 = {p \over {p+1}}.
\ee
Hence the value of $\beta$ is not integer and does not fit a Chebyshev
polynomial (\ref{Cheby}). However, the existence of a finite number
of kinks related to $k$ supports the idea. Perhaps we must
scale $\phi$ somehow. We present now a tentative argument in this line.

First, we must recall that primary fields of the form $\Phi_{(1,m)}$
or $\Phi_{(n,1)}$ are identified in the Liouville theory
with exponentials of the field corresponding to vertex
operators of the Dotsenko-Fateev construction of minimal models
\cite{Mans}.
Since the remaining primary fields
can be generated as products of fields of those two types,
we may think that the identification with the Dotsenko-Fateev
construction holds throughout.
The elementary field is then represented as
\be
\varphi \equiv
\Phi_{(2,2)} = \exp\left(i\,\alpha_{(2,2)}\,\phi \right).
\ee
This is not quite of the form (\ref{id}). Neither is it real.
It is then reasonable to take
\be
\varphi = \cos\left(\alpha_{(2,2)}\,\phi \right),  \label{exactid}
\ee
which amounts to a rescaling of $\phi$ in (\ref{id}). Now, we should
have an even integer for the quotient of the argument of (\ref{sG}) by
the argument of (\ref{exactid}),
\be
\frac{\beta}{\alpha_{(2,2)}} =
\frac{\sqrt{p \over {p+1}}}{{-1\over{2\,\sqrt{p\,(p+1)}}}} = -2\,p.
\ee
It is indeed an even integer, though negative and of absolute value
larger than we need, producing extra ground states.
It is easy to see that the required value is
$2\,(p-1)$, for example, using the Ising model with $p=3$ and
2 ground states. In fact, it is possible to obtain this precise
value. We must recall that in the Dotsenko-Fateev construction
primary fields are defined up to a conjugation. It is thus possible to
use instead of $\beta = \alpha_{(1,3)}$ its conjugate,
$\alpha_{(p-1,p-2)}$. We then have
\be
{\alpha_{(p-1,p-2)} \over \alpha_{(2,2)}} =
\frac{-{{p-1}\over{\sqrt{p\,(p+1)}}}}{{-1\over{2\,\sqrt{p\,(p+1)}}}} =
2\,(p-1),
\ee
as desired.

We believe that this argument shows that the equivalence between
both descriptions can be made rigorous. However,
we must emphasize that the sine-Gordon theory or,
in general, ATFT are exact qantum theories whereas
Landau-Ginzburg theories are classical Lagrangians.
Any discrepancy between them is to be explained
within this philosophy. So it happens for the
kinetic term, which we have neglected so far. The sine-Gordon
kinetic term is written as an infinite series expansion under
the identification (\ref{id}) or (\ref{exactid}),
\be
(\partial\phi)^2 = {1 \over {{\rm sin}^2\phi}}\,(\partial\varphi)^2 =
(1+\varphi^2+\varphi^4+\cdots)\,(\partial\varphi)^2.
\ee
All the terms but the first are irrelevant fields. Hence it is
possible to keep just the ordinary kinetic term in
the Landau-Ginzburg Lagrangian.

Field identifications similar to (\ref{id}) have in fact been
proposed in a slightly different context, that of fusion rings.
A fusion ring is the algebraic structure of primary fields of
some rational conformal field theory. It was discovered by Gepner
that the fusion rings of $SU(n)$ Wess-Zumino-Witten
models can be expressed in terms of potentials, called fusion
potentials \cite{Gepner}. Well, the fusion potentials for $SU(2)$
are the Chebyshev polynomials of the first kind.
Here also the most relevant field
is to be identified with $\cos\phi$. The other primary
fields are Chebyshev polynomials of the second kind and
reproduce the fusion rules.
We must remark that one should
not think of the fusion potential as the same as the
Landau potential. Both produce rings of perturbations (actually algebras)
by quotienting the ring of polynomials by the ideal generated by
the derivatives of $V$ (redundant fields). However, the ring
of perturbations of the fusion potential includes all primary fields
whereas that of the Landau potential only includes relevant primary fields.
Nevertheless, there is some resemblance between both structures,
as is manifested by the form (\ref{Cheby2}) found for
the renormalized field $\left[\varphi^{2\,k-2}\right]$ as
a Chebyshev polynomial of the second kind.

It is tempting
to consider the Gepner's potentials for $SU(n)$ as
a generalization of Chebyshev polynomials suitable for our purposes.
Before proceeding, we need to
recall how Gepner's potentials are constructed \cite{Gepner}.
The primary fields
of the Wess-Zumino-Witten model with Kac-Moody symmetry $SU(n)_l$
correspond to the representations of $SU(n)$ with dominant weights
up to level $l$. Gepner associates to each primary field the
character of its corresponding representation, which can be
calculated, for example, with the Weyl character formula.
The characters are functions of as many variables as the
rank of the group and are invariant under the Weyl group.
This is clearly exhibited by the Weyl formula, which involves
summation over the elements of the Weyl group.

Gepner's fusion potential is
\be
V = {1 \over m} \sum_{k=1}^n {\rm e}^{i\,m\,2\pi\,\phi_k},   \label{Gpot}
\ee
where $2\pi\phi_k$ are angular variables, constrained by $\sum\phi_k=0$,
that parametrize the Cartan subalgebra and $m$ is some integer.
Since $V$ is a completely symmetric polynomial of
$$q_k = {\rm e}^{i\,2\pi\,\phi_k},$$
it can be in turn
expressed as a polynomial of the elementary symmetric
monomials in $q_k$,
\be
\sigma_r = \sum_{i_1,\cdots,i_r} q_{i_1}\cdots q_{i_r},~~~r=1,\cdots,n-1.
\label{elfi}
\ee
As functions of $\phi_k$, these
are the characters of the fundamental representations of $SU(n)$.
They correspond to the elementary fields or order parameters.
This identification generalizes (\ref{id}).
Sice $\sigma_r$ are symmetric in $\phi_k$,
we can restrict the domain of
these variables to a fundamental region of the Weyl group.
Furthermore, as functions of $q_k$ they are periodic with $\phi_k$.
One can see that the fundamental domain of $\phi_k$ is
the $n-1$-simplex formed by the fundamental weights
with $D_n$ symmetry.
Its geometrical center is at ${\bm \phi} = {\bm \rho}/n$,
where ${\bm \rho}$ is the Weyl vector. It can be shown that
all the elementary fields (\ref{elfi}) vanish on this point.
The values of the elementary fields on the vertex of
the fundamental domain corresponding to ${\bm \omega}_k$ are
\be
\sigma_r = \left(\begin{array}{c} n\\r \end{array}\right)\,
{\rm e}^{i\,{2\pi}\,{{k\,r}\over n}},
{}~~~r=1,\cdots,n-1.
\label{sig-ver}
\ee
One can see that the mapping from ${\bm \phi}$ to
$\sigma_r$ amounts to a deformation of the simplex that preserves
its symmetry properties.

The fusion potential (\ref{Gpot}) is certainly of
the ATFT type. However, if we take $\bm\phi$ to stand for
the ATFT field, we should consider the slightly
different form
\be
V = -\sum_{k=1}^n
{\rm exp}\left({i\,m\,2\pi\,{\bm \alpha}_k\cdot{\bm \phi}}\right),
\label{Tpot}
\ee
where $\bm\alpha_k$ are the set of positive roots plus
minus the highest root $\boldmath\alpha_n \equiv \alpha_0$.
This potential possesses two types of symmetries, rotations-reflections
and translations. The first type is the symmetry of
the extended root system,
in this case $D_n$. The translational symmetry
\be
{\bm \phi \; \rightarrow \; \bm\phi} + {l\over m}\,{\bm\omega}_k,
{}~~l \in {\rm Z},~~k=1,\cdots,n-1,  \label{transl}
\ee
is due to its exponential form
and is especially interesting since it is the cause of
the existence of solitons. When $l=m$, the transformation (\ref{transl})
maps the fundamental domain of the elementary fields (\ref{elfi})
onto another domain, producing an automorphism of these fields,
\be
\sigma_r = {\rm e}^{i\,{2\pi}{r\over n}}\, \sigma_r.
\ee
It corresponds to the $Z_n$ symmetry of the $W_{(n)}^{\,p}$ models.

The AFTF potential is not real (like Gepner's)
but we can take its real part.
This real potential is able to be related with a Landau potential.
It can be easily plotted for the $W_3$
case and looks like a 2d generalization of sine-Gordon.%
\footnote{This fact was noted in \cite{DoRa}.}
There are valleys placed on a triangular array and small
mountains on a hexagonal array (Kagom\'e lattice) interlaced
with the former. (See fig. 1.)
Now, we are to compare the structura of extrema inside a
fundamental domain of $\sigma$ with the structure of extrema
of $W_{(3)}^{\,p}$ Landau potentials \cite{I}.
It has been observed before \cite{Naka}
that the ATFT with
$W_3$ symmetry in the restricted case has solitons that
match the kinks between ground states of the corresponding
IRF models of Jimbo et al. We showed in \cite{I} that
the Landau potential reproduces these ground states.
These ground states appear as a consequence of the symmetry
under translations (\ref{transl}). A translation (\ref{transl})
produces new ground states only for $l \leq m$.
Hence there must be as many as
fit in a triangle of side $m$. This is the maximum number of minima
found before for the Landau potential with $p=m+3$ \cite{I,II}.
It is also important to check that the other extrema,
maxima and saddle points, coincide as well.
It can be done by inspection on fig. 1 and comparison with
the results in \cite{I,II}.

In the case $n=3$, the generic symmetry of ATFT potentials,
$D_n$, coincides with the Weyl symmetry $S_n$ of the expression
(\ref{elfi}) for the elementary fields.
Therefore, we can express the potential
as a polynomial in the elementary fields,
$\sigma$ and $\bar\sigma$, according to Gepner's idea.
However, the polynomials thus
constructed do not have the correct structure of extrema.
This is because the minima situated on the border of the
fundamental domain
are pushed to infinity after expressing the potential in terms of
$\sigma$ and $\bar\sigma$ (The jacobian of this change of variables
vanishes on the border.) Therefore, the Landau potential misses them.
We could enlarge the fundamental domain
by shifting the border outwards up to the nearest maxima.
Unfortunately, this would spoil the Weyl symmetry of the potential,
which could not then be expressed in terms of
$\sigma$ and $\bar\sigma$.
There seems to be no way to obtain a suitable polynomial
to be identified with the Landau potential.

For $n > 3$ the structure of the extrema of
the real part of the ATFT potential in the fundamental domain of
$\sigma_r$ also agrees with the structure of the extrema of
the corresponding Landau potential.
However, now it is not even possible to express
the real part of the ATFT potential in terms of $\sigma_r$,
since the former has only $D_n$ symmetry and
the latter has full Weyl symmetry.
It is known that ATFTs only possess the symmetry of the extended
Dynkin diagram, which can be identified with a subgroup of
the Weyl group isomorphic to the semidirect product of
the center of the Lie group and complex conjugation
\cite{OlTu}. This group is $D_n$ for the Lie algebra $A_{n-1}$,
while the full Weyl group is $S_n$.
This fact may seem an undesirable feature of ATFT potentials
in regard to their relation with Landau potentials.
On the contrary, it turns out to be a necessary property for
the soliton spectrum of restricted ATFT to fit the kink structure
provided by Landau potentials; namely, the kinks that interpolate
between ground states belonging to a diagram of dominant weights.
The kinks that actually correspond to solitons (indecomposable into
others more elementary) must have weights belonging to some fundamental
representation as topological charges. However, not all the weights
of each fundamental representation can appear, as will be seen in
a concrete example below.

It is not difficult to show for $n > 3$
that the ground state structure of the restricted ATFT potential
is the same as that of the Landau potential.
We shall do it for $n=4$, with potential%
\footnote{We omit the coupling constant $\beta$
(or number $m$ in the restricted case).}
\be
V({\bm\phi}) =  -
\sum_{k=0}^3 {\rm exp}\left({i\,2\pi\,{\bm\alpha}_k\cdot{\bm\phi}}\right).
\ee
This potential is real along the directions of the weights
of the fundamental representations and the horizontal roots
of $SU(4)$. For the directions along the fundamental weight of
the representation 4, ${\bm\omega_1}$,
$$
{\bm\phi} = \phi\,{{\bm\omega_1}\over |{\bm\omega_1}|} \; \Rightarrow \;
{\bm\alpha}_2\cdot{\bm\phi} =  {\bm\alpha}_3\cdot{\bm\phi} = 0,
$$
and along the fundamental weight of $\bar 4$, ${\bm\omega_3}$,
$$
{\bm\phi} = \phi\,{{\bm\omega_3}\over |{\bm\omega_3}|} \; \Rightarrow \;
{\bm\alpha}_2\cdot{\bm\phi} = {\bm\alpha}_1\cdot{\bm\phi} = 0,
$$
the potential is
\be
V({\bm\phi}) = - 2 - 2 \cos{2\pi\,\phi\over{{\sqrt 3}/2}}.
\ee
Due to the $D_4$ symmetry,
this form of the potential holds for the other weights of
the representations 4 and $\bar 4$, completing the diagonals
of a square.

Along the fundamental weight of the representation 6,
${\bm\omega_2}$, the potential is
\be
V({\bm\phi}) = - 2 - 2\,\cos\left(2\pi\,\phi\right).
\ee
We know again that this form holds for those weights of
the representation 6 related to ${\bm\omega_2}$
by the $D_4$ symmetry, namely, the horizontal weights.
However, the potential along the two vertical weights is
now different. Since
$$
{\bm\alpha}_1\cdot{\bm\phi} = {\bm\alpha}_3\cdot{\bm\phi} =
-{\bm\alpha}_2\cdot{\bm\phi} = -{\bm\alpha}_0\cdot{\bm\phi}
$$
the potential is
\be
V({\bm\phi}) = - 4 \cos(2\pi\,\phi).
\ee
A similar form is found for the horizontal roots.
With these results it is easy to figure out the
overall structure of the potential.
There are minima ($V=-4$) situated on a body centered
cubic (bcc) lattice and maxima ($V=4$) situated on
a similar lattice interlaced with the previous one (fig. 2).
Let us remark once more the absence of Weyl symmetry.
It is reflected in the fact that the potential
for the two vertical weights of 6 is different from that
for the four horizontal ones. In particular, the solitons
that link the minima corresponding to
the two vertical weights of 6 go over a higher wall
(twice as high)
and are therefore more energetic than those that link
the minima corresponding to the horizontal weights of 6.

Let us see what happens in the restricted case.
The kinks corresponding to the two vertical weights of 6
are in the spectrum above those corresponding to
the horizontal ones. Thus the simplest model contains the
kinks associated to the weights of 4 and $\bar 4$ and only
the horizontal weights of 6.%
\footnote{This fact has already been noted in the context of
the computation of exact soliton solutions \cite{Mc2}.}
In other words, we have only the kinks associated to
the dominant weight diagram at lowest level.
This diagram occurs in the fundamental domain of $\sigma_r$ when $m=1$.
We further find in this domain one maximum at the center and
six saddle points situated between minima.
This is the ground state structure
given by the Landau potential of $W_{(4)}^{\,5}$ \cite{I}. The next
restricted model
allows double kinks in the directions of 4 and $\bar 4$ and
in the horizontal of 6. It allows as well the single kink (and antikink)
in the vertical direction of 6.
They all fit in the dominant weight diagram at the next level.
Once more, one can see that also the maxima and saddle points
correspond to the Landau potential of $W_{(4)}^{\,6}$.

\section{Conclusions}

The role of Landau-Ginzburg and ATFT Lagrangians as classical
descriptions of deformed 2d CFT has been studied
to find out to what extent they are equivalent. The first essential
observation is that it is necessary to consider quantum corrections
to the Landau-Ginzburg Lagrangian to reproduce the multiple
phase coexistence given by the ATFT at the outset.
The lowest order corrections
for the tricritical Ising model
($M_4$) have been obtained and shown to produce the
adequate type of renormalization.

For more complicated models of the minimal Virasoro series,
it has been argued that the renormalized potential
coincides with a Chebyshev polynomial of the first kind.
Since this polynomial is formally the fusion potential
obtained by Gepner for $SU(2)$, we have studied whether
Gepner's construction of
potentials for $SU(n)$ can be adapted to ATFT.
We have seen that his identification of elementary LG fields
in terms of (the Toda field) $\bm\phi$ as
Lie group characters (\ref{elfi})
is adequate with regard to their expected symmetry properties.
We have also shown that the fundamental domain of definition
of elementary fields contains the correct structure of minima
according to the spectrum of solitons in restricted ATFT.
Furthermore, all the extrema (minima, maxima and saddle points)
agree with those obtained from Landau potentials.

Unfortunately, it is crucial in Gepner's
construction that the potential be a completely symmetric
function of $\phi_i$ or, in other words, that it have
Weyl symmetry. This condition is not satisfied by the ATFT
potential. Moreover, in the $W_{(3)}$ case, in which it is
indeed satisfied, even though a Landau like potential
can be obtained by Gepner's method, it is not the correct
Landau potential, known from \cite{I}:
It misses the ground states on the boundary.
We can therefore conclude that the relation between both
types of classical Lagrangians is indirect:
They yield the same extrema
but cannot be related by a change of variables.

\vspace{8mm}
\noindent
{\large \bf Acknowledgements}\\[2mm]
\nopagebreak
I am grateful to F. Alexander Bais and C\'esar G\'omez
for conversations.

\newpage
\def\epsfsize#1#2{1.0#1}
\centerline{\epsfbox{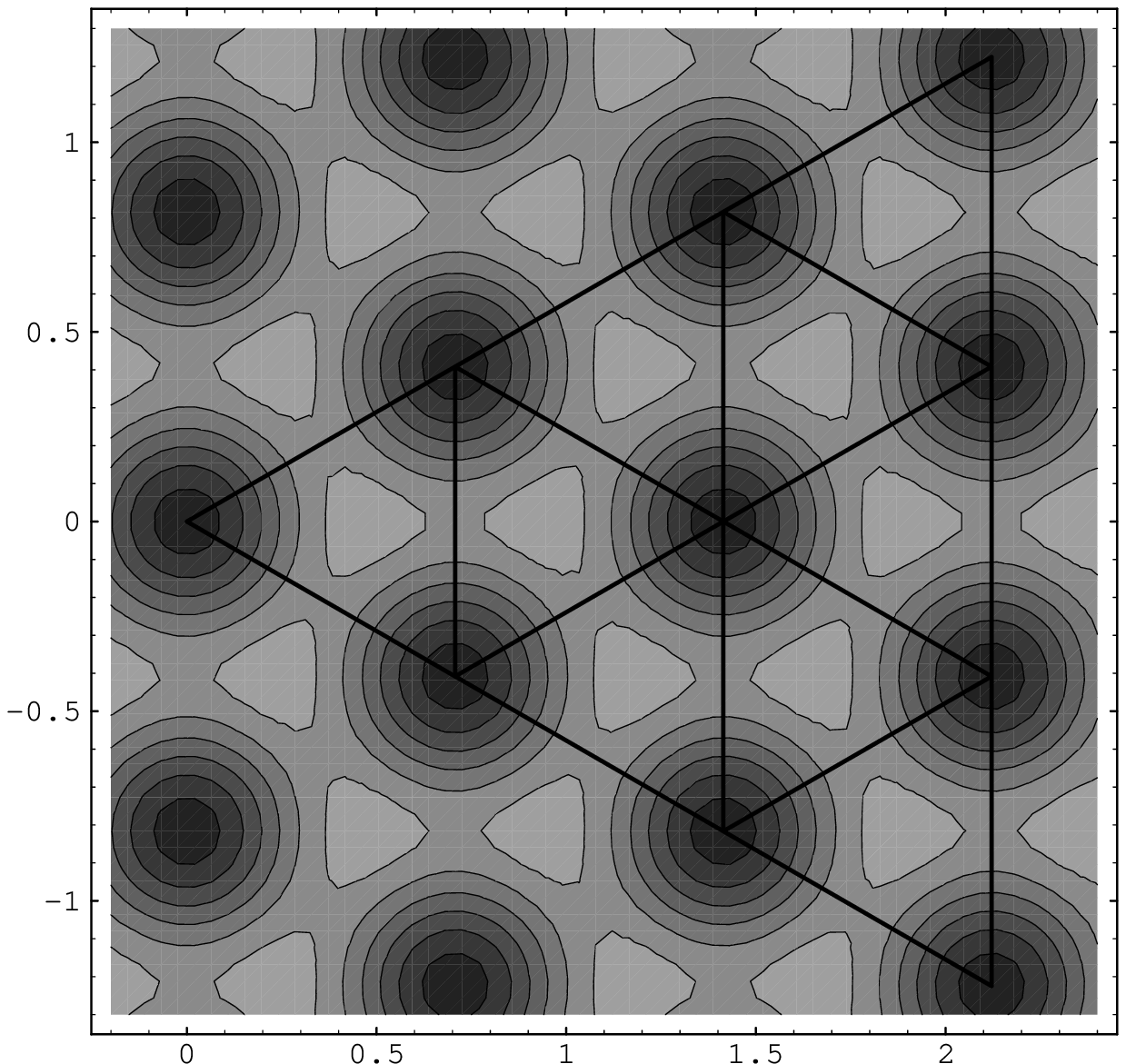}}
\vskip 2cm
\begin{center}
{\bf Fig. 1}\hspace{3mm} Structure of extrema (contour plot)
for the real part of the
$n=3$ ATFT potential with the dominant weight diagram of level 3.
\end{center}

\newpage
\def\epsfsize#1#2{1.0#1}
\centerline{\epsfbox{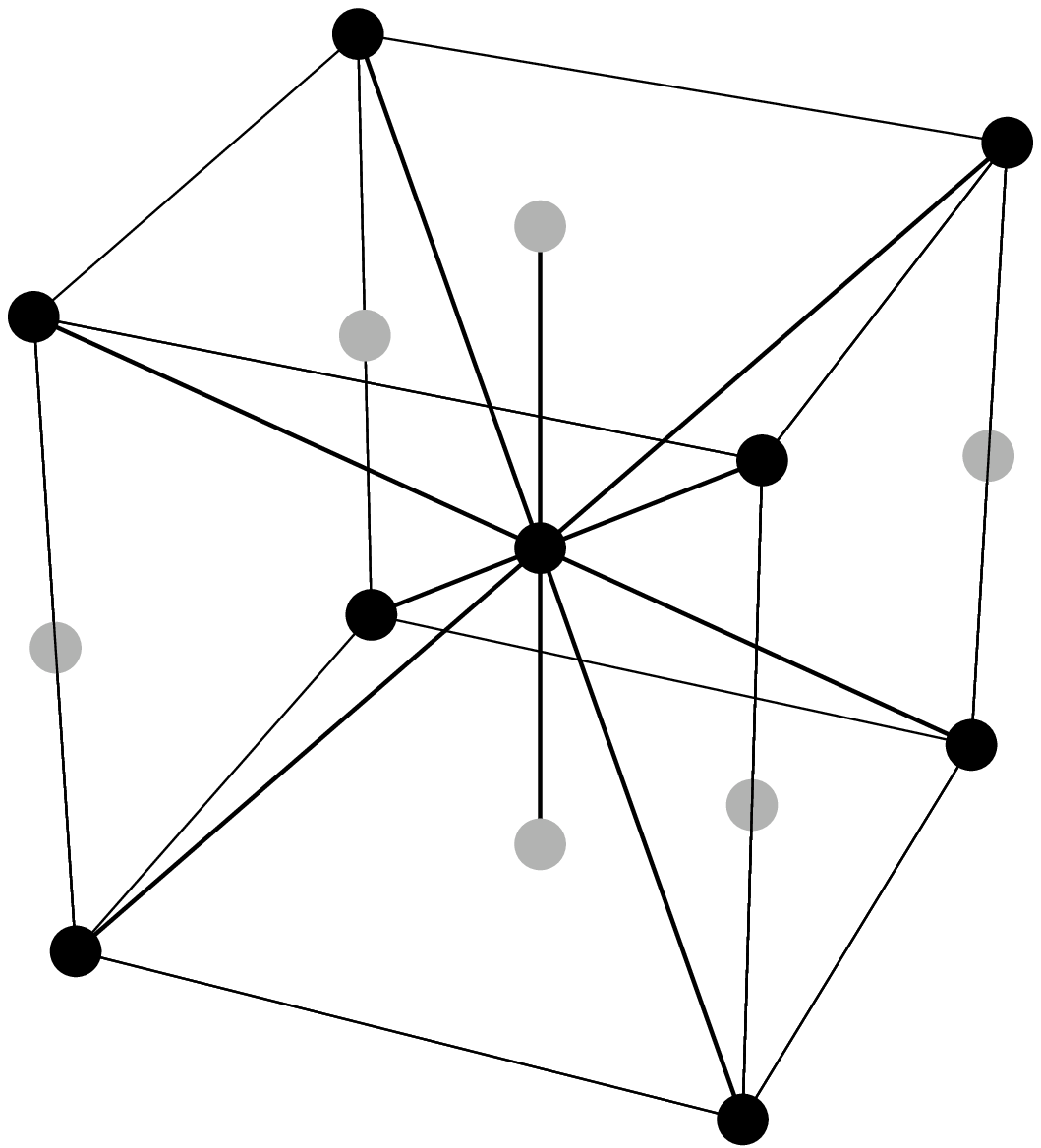}}
\vskip 2cm
\begin{center}
{\bf Fig. 2}\hspace{3mm} Unit cell of the lattice of extrema for
the real part of the $n=4$ ATFT potential.
Minima $V=-4$ are displayed in black
and maxima $V=4$ in grey.
\end{center}

\end{document}